\documentclass[conference]{IEEEtran}
\IEEEoverridecommandlockouts
\usepackage{cite}
\usepackage{amsmath,amssymb,amsfonts}
\usepackage{algorithmic}
\usepackage{graphicx}
\usepackage{float}
\usepackage{subfigure}
\usepackage{textcomp}
\usepackage{xcolor}
\def\BibTeX{{\rm B\kern-.05em{\sc i\kern-.025em b}\kern-.08em
    T\kern-.1667em\lower.7ex\hbox{E}\kern-.125emX}}
\begin{document}

\title{VSA: Reconfigurable Vectorwise Spiking Neural Network Accelerator}

\author{\IEEEauthorblockN{Hong-Han Lien$^1$, Chung-Wei Hsu$^2$, and Tian-Sheuan Chang$^2$, \textit{Senior Member, IEEE}}
\IEEEauthorblockA{\textit{$^1$AI Graduate Program, $^2$Institute of Electronics, Nat'l Chiao Tung Univ.}, Taiwan}
\thanks{H. -H. Lien, C. -W. Hsu and T. -S. Chang, "VSA: Reconfigurable Vectorwise Spiking Neural Network Accelerator," 2021 IEEE International Symposium on Circuits and Systems (ISCAS), 2021, pp. 1-5, doi: 10.1109/ISCAS51556.2021.9401181.}
}
\maketitle

\begin{abstract}
Spiking neural networks (SNNs) that enable low-power design on edge devices have recently attracted significant research. However, the temporal characteristic of SNNs causes high latency, high bandwidth and high energy consumption for the hardware. In this work, we propose a binary weight spiking model with IF-based Batch Normalization for small time steps and low hardware cost when direct training with input encoding layer and spatio-temporal back propagation (STBP). In addition, we propose a vectorwise hardware accelerator that is reconfigurable for different models, inference time steps and even supports the encoding layer to receive multi-bit input. The required memory bandwidth is further reduced by two-layer fusion mechanism. The implementation result shows competitive accuracy on the MNIST and CIFAR-10 datasets with only 8 time steps, and achieves power efficiency of 25.9 TOPS/W. 
\end{abstract}


\section{Introduction}
Artificial neural network (ANN) becomes popular in recent years due to its excellent performance, but it contains a large amount of data transfer and complex calculations that is not suitable for resource-constrained battery-powered mobile devices. Therefore, spiking neural network (SNN), a brain-inspired computing models, is currently developed as an alternate to ANN since its binary spikes transmission is relatively simple for hardware implementation. So far,~\cite{yin2017algorithm,frenkel2019morphic,park20197,chuang202090nm} have demonstrated their application specific designs enable inference or learning with extremely low power consumption, but they are limited to small neural network architectures or simple datasets. Only the large-scale neuromorphic hardware systems such as TrueNorth~\cite{akopyan2015truenorth} and Loihi~\cite{davies2018loihi} can perform the large-scale neural network and can reconfigure. However these modern SNN architectures achieve lower throughput and higher energy per neuron, compared to ANN accelerators.~\cite{narayanan2020spinalflow} attribute this phenomenon to the temporal aspect in SNN that process data across many time steps. This makes the inference time extremely long and repeatedly accessing the same data from memory will consume a lot of unnecessary energy. In addition, a stateful spiking neuron needs to record membrane potential every time step that will increase off-chip memory transfer, which leads to more power consumption.

To address above issues, this paper proposes a reconfigurable vectorwise SNN hardware accelerator that needs small time steps and low hardware cost and memory bandwidth through model and hardware optimizations. For the SNN model, this paper integrate the binary neural network (BNN) training method into SNN to directly train a binary weight spiking model, which minimizes the off-chip memory transfer and on-chip storage. The first layer is designed to be an encoding layer~\cite{wu2019direct}, which receives multi-bit positive inputs and generate spikes for the next layer. This can greatly reduces the time steps required for inference. Furthermore, the proposed integrate and fire (IF)-based Batch Normalization (BN) reduce BN overhead on hardware. For the hardware part, this design can reconfigure to support different inference time steps and models, and also supports the encoding layer to receive multi-bit inputs as well as spiking layer processing with spike inputs with a vectorwise input format. In addition, the memory bandwidth is further reduced by tick batching~\cite{narayanan2020spinalflow} and layer fusion mechanism. The result shows better performance and implementation cost compared to other approaches.

\section{Binary Weight Spiking Neural Network}

\begin{figure}[t]
\centering
\subfigure[The behavior of IF]{
\label{IF activity}
\includegraphics[width=0.15\textwidth]{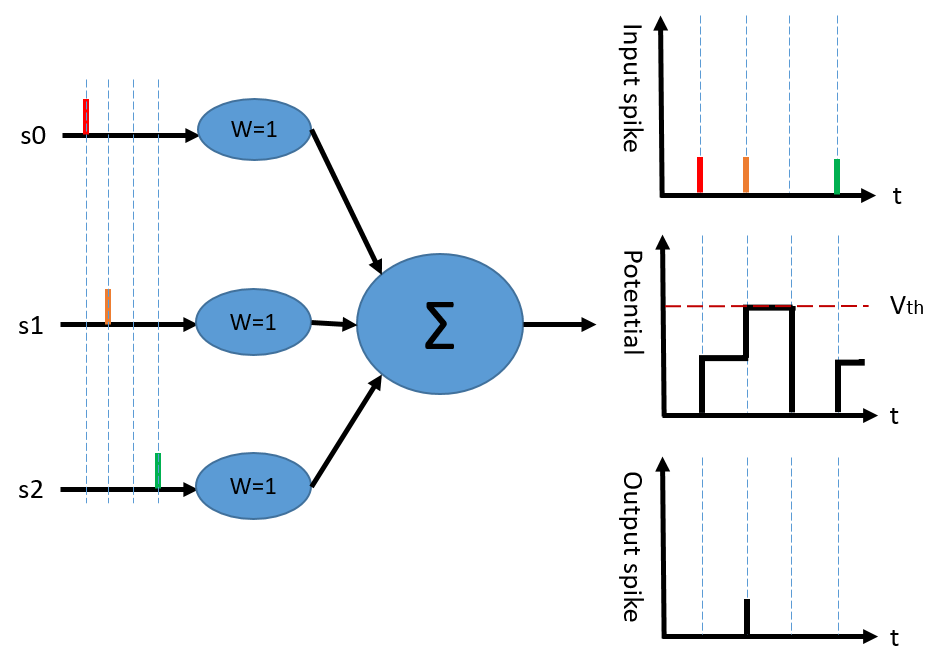}}
\subfigure[IF architecture]{
\label{IF neuron architecture}
\includegraphics[width=0.3\textwidth]{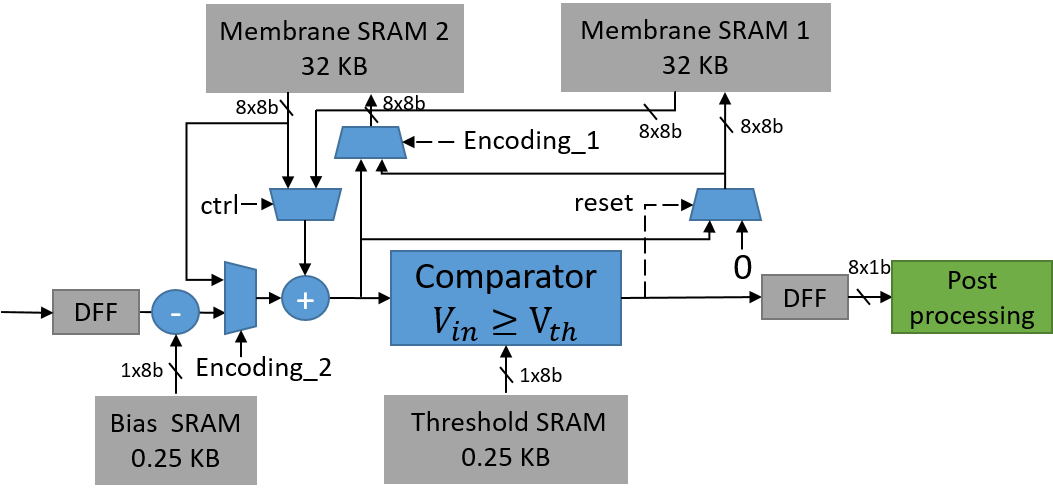}}
\caption{IF neuron}
\label{IF neuron}
\end{figure}

\subsection{Spiking Networks}
This work adopts spatio-temporal back propagation (STBP)~\cite{wu2018spatio} to directly train the network and replace Leaky Integrate-and-Fire neuron to Integrate-and-Fire neuron (IF) to reduce hardware cost.
Fig.~\ref{IF activity} shows the behavior of IF neurons, which includes membrane potential update (Eq. (1)) and spike firing condition (Eq. (2)). 
\begin{align}
    V^{l}[t+1]&=V^{l}[t](1-o^{l}[t])+w^{l}o^{l-1}[t+1]\\
    o^{l}[t+1]&=f(V^{l}[t+1]), 
    f(x)
    \left\{
             \begin{array}{lr}
             1, & x \geq V_{th}  \\
             0, & \text{otherwise}.
             \end{array}
    \right.
\end{align}
where $V^{l}[t]$ denotes the membrane potential at the time step t, $w^{l}$ is the weights, $o^{l-1}[t+1]$ is the output spike from previous layer at the time step t+1, f(.) is the firing function and $V_{th}$ is the threshold. During model inference, neurons receive asynchronous weighted input and accumulate them into membrane potential. Whenever membrane potential reaches the threshold, neurons fire an output spike and reset membrane potential.

\subsection{IF-based Batch Normalization}
To further minimize off-chip memory transfer and on-chip storage, we further combining the Binary Neural Networks~\cite{hubara2016binarized} training method into the above method to reduce the model weights to (-1,+1). For binary neural network, batch normalization~\cite{ioffe2015batch} is a widely mechanism to enable training of the binary weight networks. However, BN suffers from complex computation and high hardware cost. To solve this problem, we propose an IF-based BN, which is inspired by the bias-based BN~\cite{yonekawa2017chip}. The IF-based BN integrates BN into IF neuron computation by rearranging the membrane potential integration and threshold in the firing condition from the original formulation as Eq. (3) to Eq. (4).
\begin{align}
    & [\gamma (\frac{x[1]-\mu}{\sqrt{\sigma^2}})+\beta]+[\gamma (\frac{x[2]-\mu}{\sqrt{\sigma^2}})+\beta]+\cdots \geq V_{th} \\
    & \sum_{t=1}^{T} [x[t]-(\mu-\frac{\sigma}{\gamma}\beta)] \geq \frac{\sigma}{\gamma}V_{th} 
\end{align}
where $\gamma, \beta, \mu, \sigma^2$ denote the BN parameters: gamma, beta, running mean, running variance and x[t] denote convolution output at time step t . As illustration in Eq. (4), the new formulation just needs to subtract the convolution output with a new bias  $(\mu-\frac{\sigma}{\gamma}\beta)$ and compare with a new threshold $\frac{\sigma}{\gamma}V_{th}$ to generate the desired output spike, which reduces the extra hardware BN overhead.

\begin{figure}[t]
\centering
\includegraphics[width=0.47\textwidth]{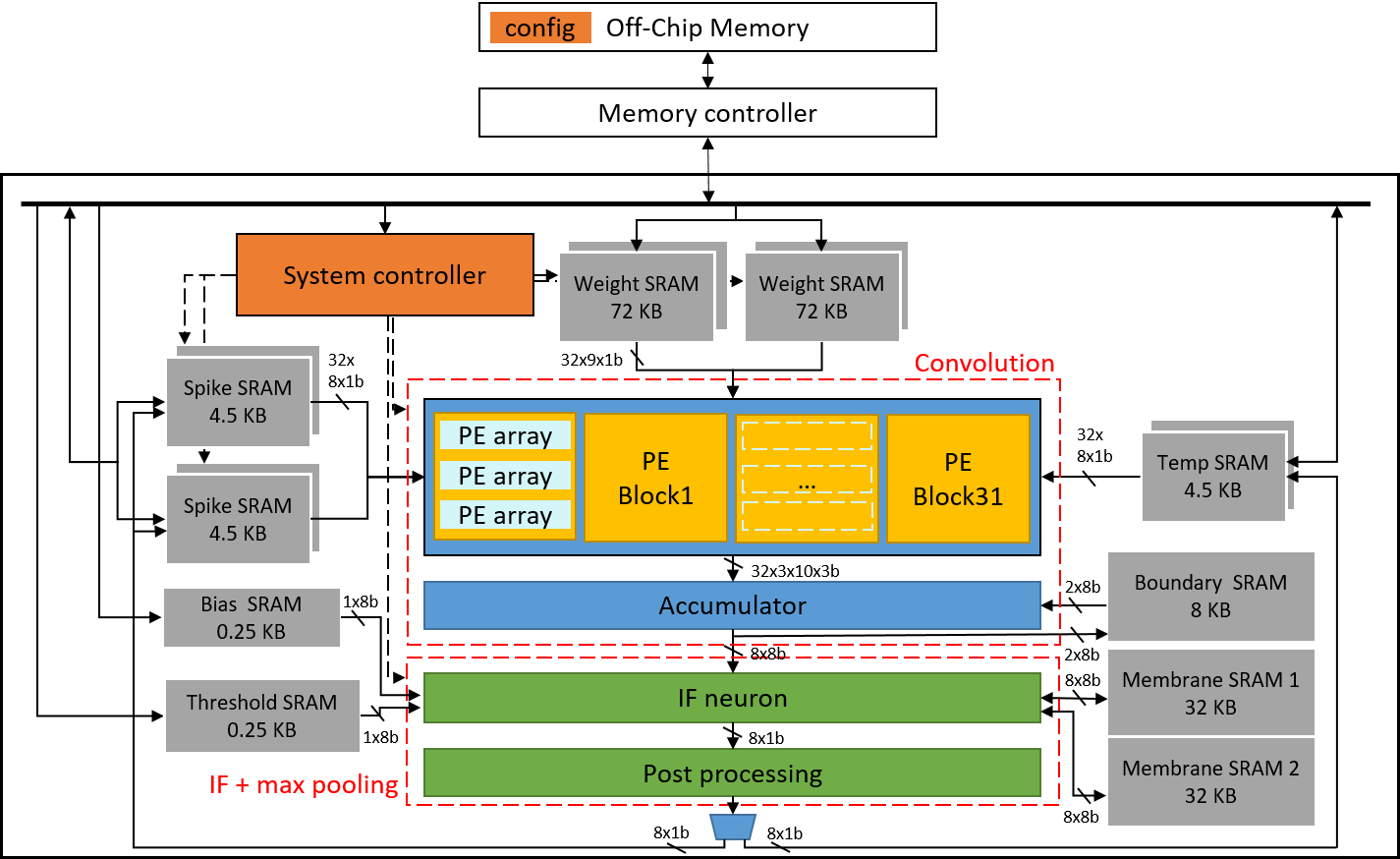}
\caption{System architecture}
\label{System architecture}
\end{figure}

\section{Architecture and Data Flow}

\subsection{Overview}

Fig.~\ref{System architecture} shows the proposed system architecture. This design accesses weight and input from off-chip memory and stores them into weight and spike SRAM buffers. For the spike ping-pong buffer, one will be accessed at time-step t and the other will be accessed at time-step t+1. The weight ping-pong buffer stores different layer's weight in each buffer. Each PE block processes one channel of input spike at time step t and each PE array computes one vector of them to generate the convolutional partial sum. Total 32 PE blocks can process 32 channels of input. The output of PEs will be accumulated by the accumulator to complete a convolution operation. Then, the IF neurons accumulate membrane potential and fire output spike accordingly. The result will go through the post processing if needed. The result will be saved to temp SRAM and then transferred to off-chip memory. Above process is repeatedly for all time steps of a layer input spike before moving to the next layer to prevent membrane potential from being transferred off and back on chip. With this, the weight data is also reused during the spike processing of all time steps. 

\begin{figure}[t]
\centering
\includegraphics[width=0.35\textwidth]{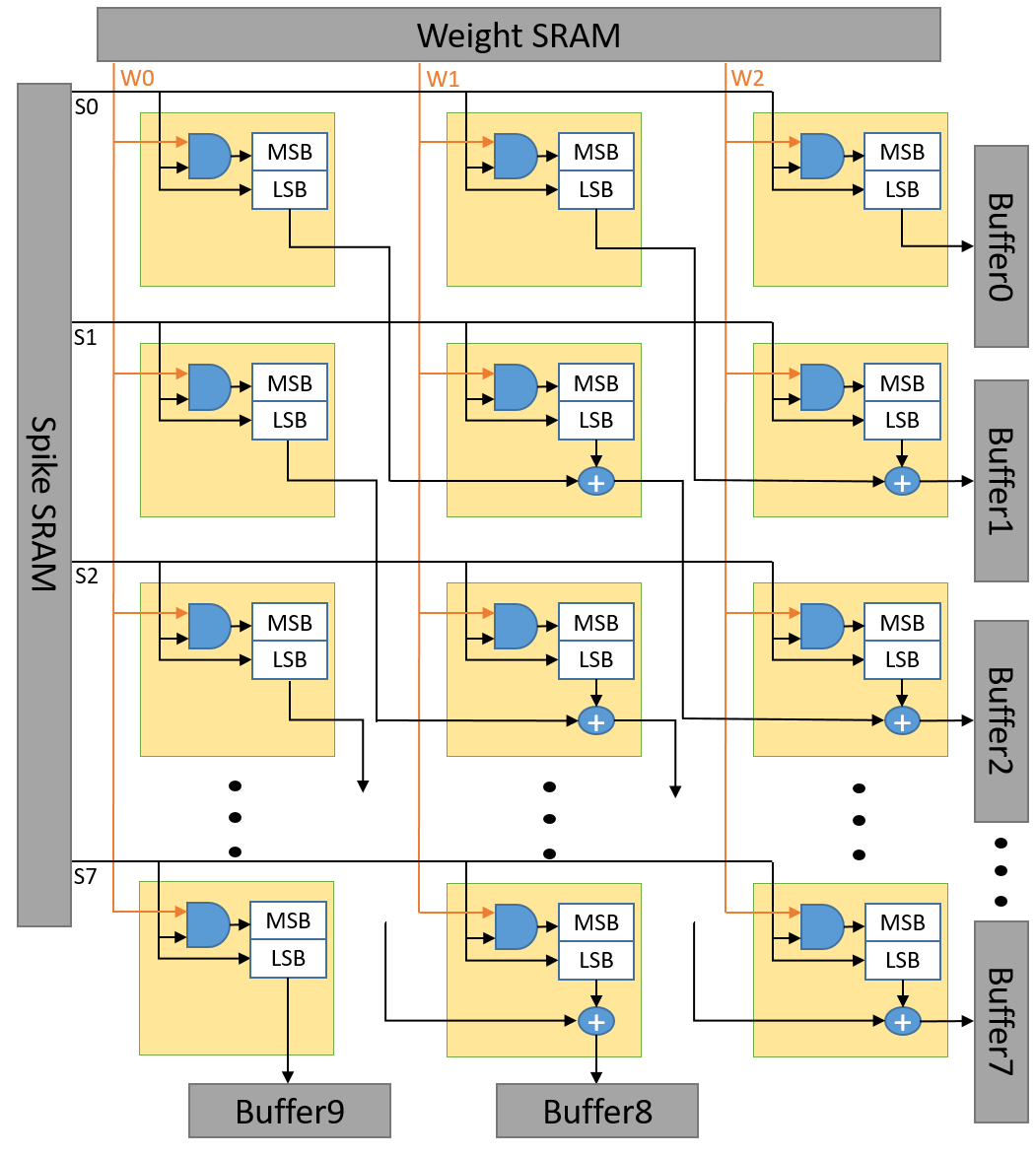}
\caption{PE array}
\label{PE array}
\end{figure}

\subsection{PE Blocks}
PE block consists of three PE arrays, which is constructed by 8$\times$3 PEs as shown in Fig.~\ref{PE array}. In the figure, eight input spikes broadcast horizontally and three weights broadcast vertically. They are multiplied using AND gate then summed up along the diagonal direction. The outputs will be stored in ten registers individually, which are partial sums of one filter column for one input channel column. 

Both spike (0,1) and weight (-1,+1) are binary value, multiplying them will get a set of possible multiplication result (-1,0,1). We represent weight using only one sign bit, so weight -1 is stored as 1 and weight +1 is stored as 0. As a result, the multiplication result can be simply computed by a logic equation $o=\left \{ s\&w,s \right \}$.

\begin{figure}[t]
\centering
\includegraphics[width=0.45\textwidth]{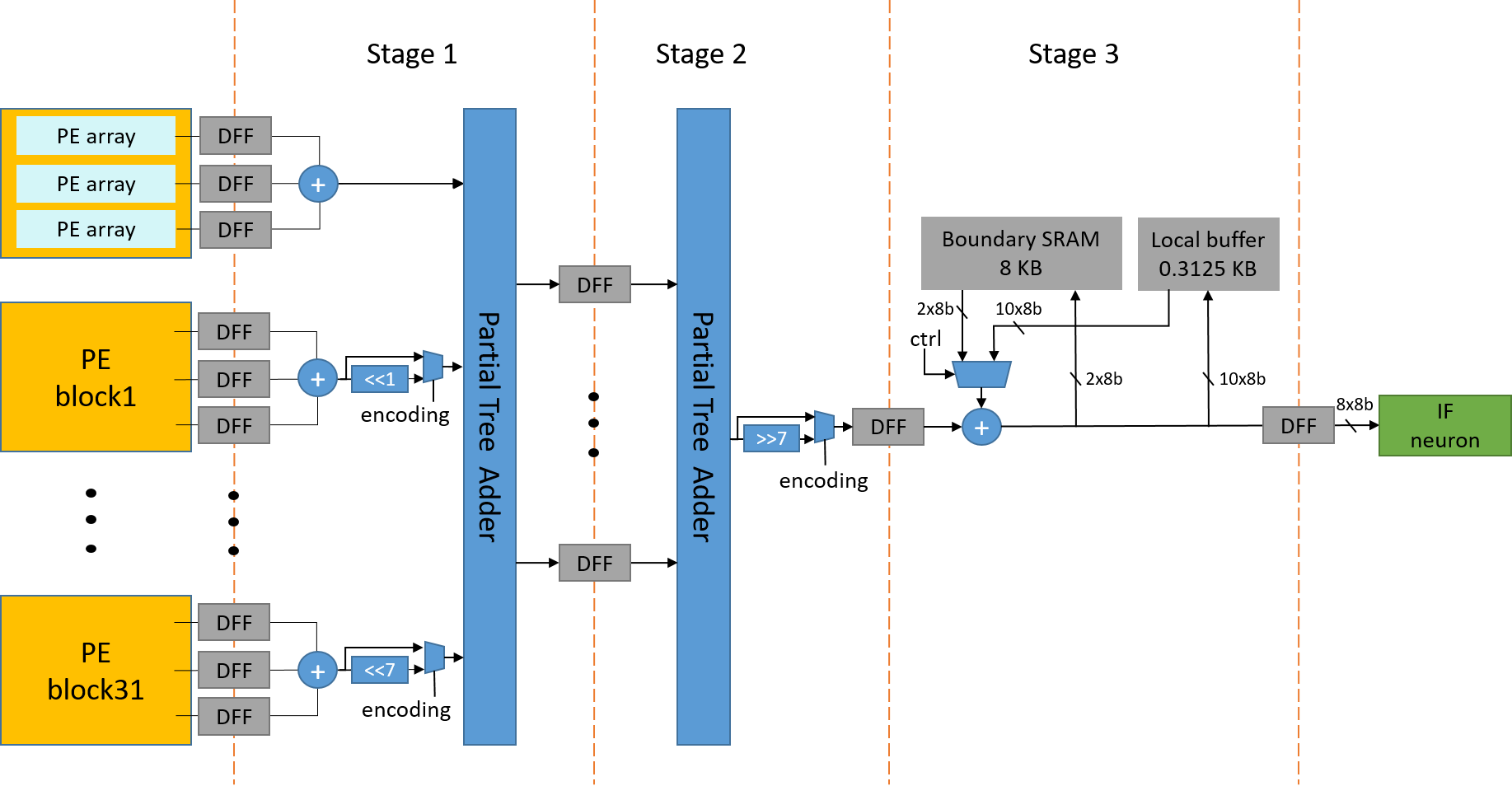}
\caption{Accumulator}
\label{Accumulator}
\end{figure}

\subsection{Accumulator}
Fig.~\ref{Accumulator} shows the accumulator, which is divided to a three-stage pipeline for shorter critical path. Three vectorwise partial sums of the same channel from three PE arrays is summed up first, and then the 32 output channels from 32 PE blocks is summed up with a tree adder that is divided to two partial tree adders. 

If input channel number is larger than 32, they will be further divided to groups with group size 32 and then each group will be sent to PEs sequentially for processing. These partial group results are then accumulated at the last stage of the accumulator to generate the final convolution outputs. Simultaneously, we will also store the boundary information of a tile into the boundary SRAM and accumulate them with output from the neighboring tile afterward.

\begin{figure}[t]
\centering
\subfigure[Data flow chart of a PE block]{
\label{Data flow chart}
\includegraphics[width=0.48\textwidth]{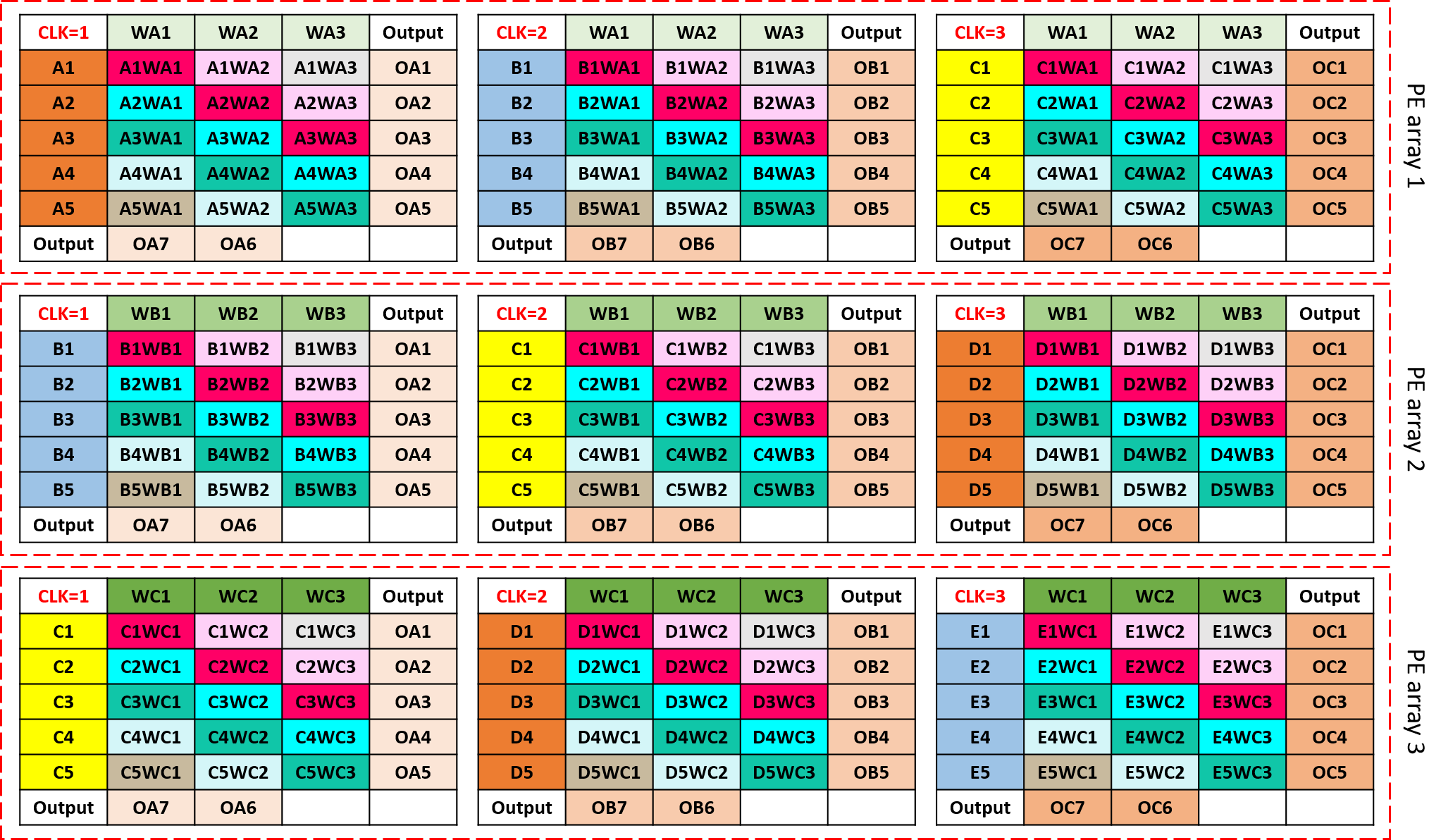}}
\subfigure[Data flow Scheduling]{
\label{Data flow Scheduling}
\includegraphics[width=0.2\textwidth]{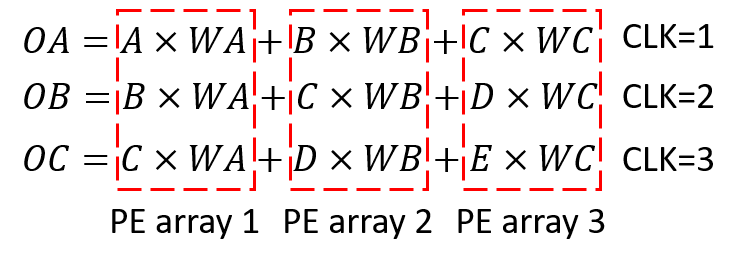}}
\caption{(a) Data flow chart for example in Fig.~\ref{conv example}. For the I/O part, the same color blocks corresponds to the same SRAM. For the computation part, the same color blocks will be summed up as partial sums. (b) Data flow scheduling expressed by mathematical formulas.}
\label{Data flow}
\end{figure}

\begin{figure}[t]
\centering
\includegraphics[width=0.35\textwidth]{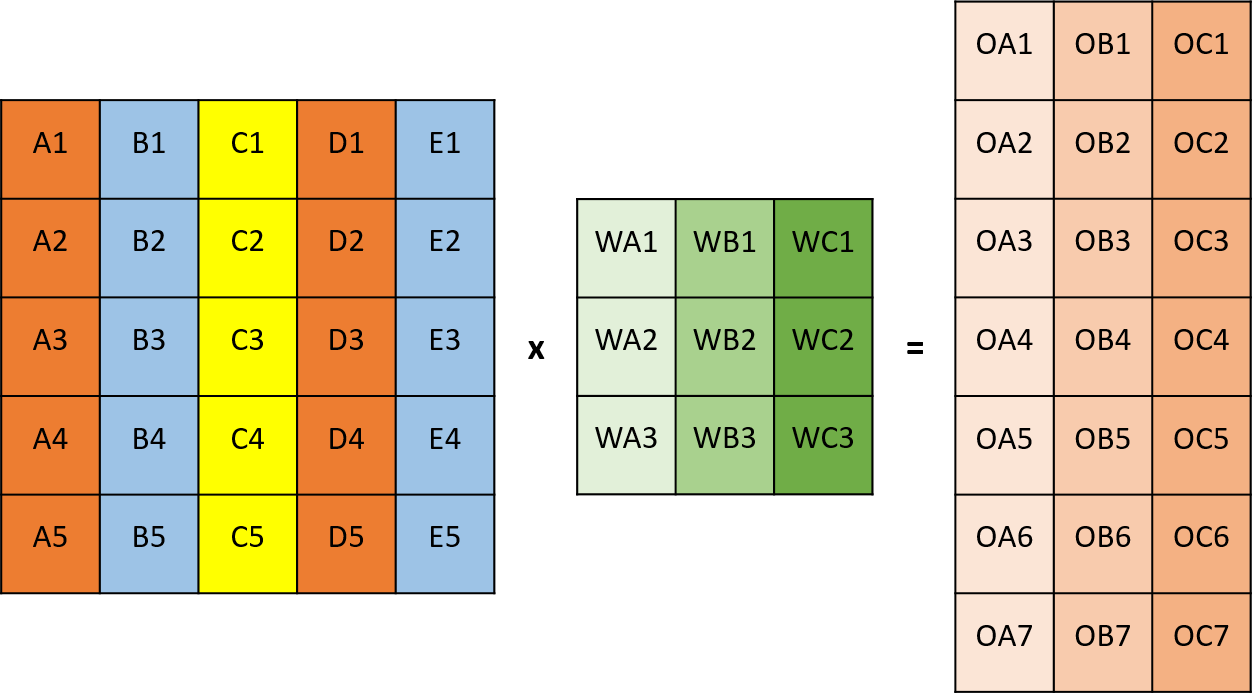}
\caption{A convolution example with 5$\times$5 inputs, 3$\times$3 weights, and 7$\times$3 outputs.}
\label{conv example}
\end{figure}

\subsection{Data flow of convolution}
Fig.~\ref{Data flow chart} shows the data flow of a PE block process on example as shown in Fig.~\ref{conv example}. This case only assume 5$\times$3 PE arrays for clarity. In our data flow, one column vector of input is broadcasted horizontally along the PE array row and one column vector of filter weight is broadcasted vertically along the PE array column. With this data flow, the products along the diagonal direction will be summed up and then accumulate with other PE arrays' outputs in the same cycle to generate convolutional results. The bottom boundary part of outputs (e.g. OA6, OA7) will be stored in the boundary SRAM to be accumulated with the top boundary part of outputs (e.g. OA1, OA2) of the neighboring tile afterward. In this schedule, all PEs are activated and contributed to the results, which achieves full hardware utilization.It is worth noting that each PE block only accesses one vectorwise input at a time that efficiently reuse the data.


This vectorwise scheduling can be expressed by mathematical formulas as in Fig.~\ref{Data flow Scheduling} that takes only three cycles to complete the example. This data flow achieve high parallelism and simple control while maintaining full hardware utilization.

\begin{figure}[t]
\centering
\includegraphics[width=0.25\textwidth]{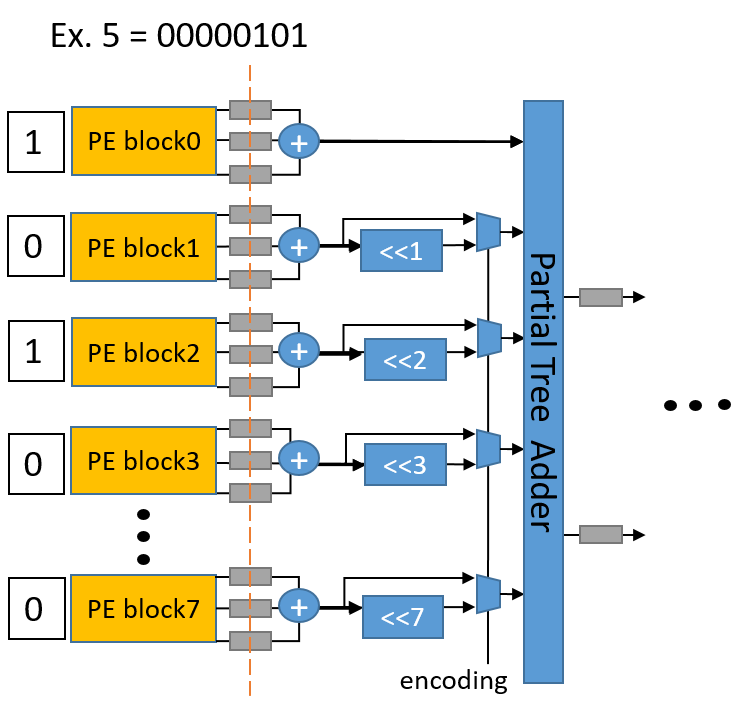}
\caption{The mapping example of encoding layer}
\label{hybrid precision example}
\end{figure}

\subsection{Encoding layer}

The first layer of our model is a encoding layer~\cite{wu2019direct} that receives multi-bit inputs and generates spikes for the following spiking layer. To support the encoding layer with the same hardware without large overhead, we split the 8-bit inputs to eight 1-bit input bitplanes, and assign each bitplane to one PE block. Eight bitplanes will takes eight PE blocks. Thus, every eight PEs block correspond to the same weight. With this, we use the first stage accumulator that shifts the PE result and sums all these bitplane results to achieve the multi-bits convolution as shown in Fig.~\ref{hybrid precision example}. 

This mechanism is applicable when the inputs are positive, thus the inputs are normalized to (0, 1) during training.

\subsection{IF Neuron}
Fig.~\ref{IF neuron architecture} shows the architecture of the IF neuron that receives the convolutional results and accumulate them with the residue membrane potential stored in SRAM. The accumulated potential is compared with the threshold, which will fire a spike and reset the membrane potential in the corresponding position once reaching the threshold.

For the encoding layer, the convolutional results will be stored to the second membrane SRAM first and then accumulated with the residue potential from the fist membrane SRAM to continuously generate output of different time steps. 


\begin{table}[t]
\caption{Two network structures on two different datasets.}
\begin{tabular}{|l|l|}
\hline
Datasets & Network structure                                                                                                                                                        \\ \hline
MNIST    & 64Conv(encoding)-MP2-64Conv-MP2-128fc-10fc                                                                                                                             \\ \hline
CIFAR-10 & \begin{tabular}[c]{@{}l@{}}128Conv(encoding)-128Conv-128Conv-MP2-192Conv-\\ 192Conv-192Conv-192Conv-MP2-256Conv-256Conv-\\ 256Conv-256Conv-MP2-256fc-10fc\end{tabular} \\ \hline
\end{tabular}
\label{network structure}
\end{table}

\subsection{Layer fusion for lower memory bandwidth}
Due to temporal input nature of SNN and limited on-chip buffer size, a naive SNN implementation will execute the model layer by layer that stores per layer output to external memory and load it back for next layer processing. This will need to save spikes of all time-steps to off-chip memory for consecutive processing and results in high data access and power consumption. To reduce this overhead, we adopt the layer fusion~\cite{alwani2016fused} that executes two layers successively inside the chip and then stores output to external. To support this, we increase the weight SRAM large enough to store the weights of two layers. The output spikes in temp SRAM won't be transferred to off-chip memory and will be directly sent to PE blocks to process the next layer's computation. The membrane potential will be stored in the second membrane SRAM and the output will replace the value in the spike SRAM that just been processed, then transfer to off-chip memory. Thus, the input and output transfer reduced by half.

\begin{figure}[t]
\centering
\includegraphics[width=0.45\textwidth]{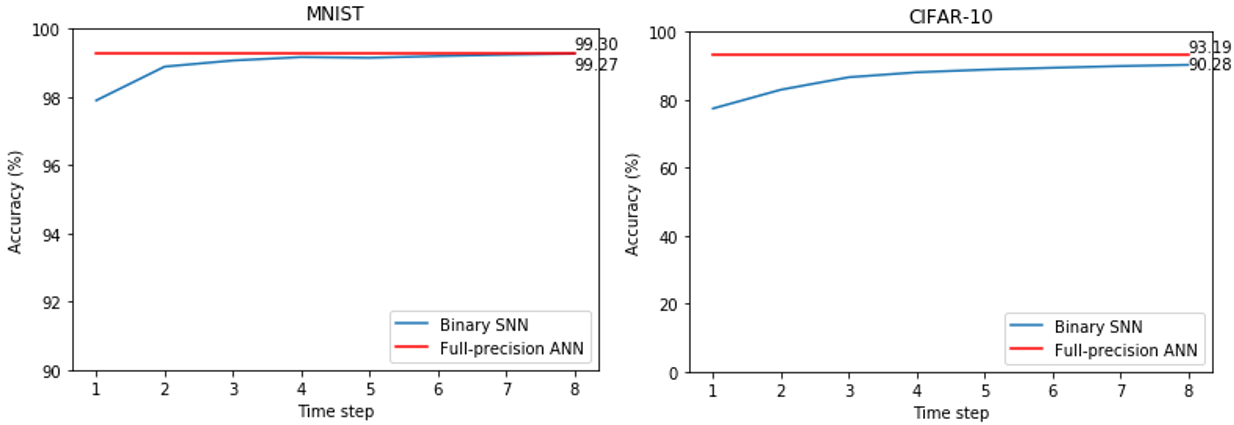}
\caption{Accuracy comparison between ANN and SNN}
\label{accuracy comparision between ANN and SNN}
\end{figure}

\section{Experimental Results}

\subsection{Network simulation result}
Table.~\ref{network structure} shows two network structures for MNIST and CIFAR-10 datasets to evaluate accuracy and analyze hardware performance. Fig.~\ref{accuracy comparision between ANN and SNN} shows the accuracy comparison between full-precision ANN model and binary weight SNN model with different time steps. Our SNN model can achieve similar accuracy as the ANN model in short time steps. Table.~\ref{comparison with others} compares existing state-of-the-art results with our model on the CIFAR-10 dataset. Our model has competitive accuracy but with significant time step reduction and binary weight, which not only avoids the huge computational cost, but also reduces the off-chip access amount.

\begin{table}[t]
\caption{Comparison with the results on CIFAR-10 of other approaches.}
\centering
\begin{tabular}{cccc}
\hline
Model                                                            & Precision      & Time steps & Accuracy \\ \hline
Sengupta et al.~\cite{sengupta2019going} & full-precision & 2500       & 91.55\%  \\
Wu et al.~\cite{wu2019direct}       & full-precision & 12         & 90.53\%  \\
Rathi et al.~\cite{rathi2020enabling}    & full-precison  & 200        & 92.02\%  \\
RMP-SNN~\cite{han2020rmp}         & full-precision & 256        & 93.04\%  \\
Wang et al.~\cite{wang2020deep}     & binary         & 100        & 90.19\%  \\ \hline
Ours                                                             & binary         & 8          & 90.28\%  \\ \hline
\end{tabular}
\label{comparison with others}
\end{table}

\begin{table}[t]
\caption{Performance summary and comparison with other designs.}
\begin{tabular}{|c|c|c|c|}
\hline
                        & This work                                           & SpinalFlow~\cite{narayanan2020spinalflow}                                        & BW-SNN~\cite{chuang202090nm}                                                  \\ \hline
Technology              & 40nm                                                & 28nm                                              & 90nm                                                    \\ \hline
Voltage (V)             & 0.9                                                 & -                                                 & 0.6                                                     \\ \hline
Frequency (MHz)         & 500                                                 & 200                                               & 10                                                      \\ \hline
Reconfigurable          & Yes                                                 & Yes                                               & fixed 5-CONV                                                 \\ \hline
Precision (bits)        & binary                                              & 8 fixed                                           & binary                                                  \\ \hline
PE number               & 2304                                                & 128                                               & 8208                                                    \\ \hline
SRAM (KB)               & 230.3125                                            & 585                                               & 12.75                                                   \\ \hline
Peak Throught (GOPS)    & 2304                                                & 51.2                                              & 64.46                                                   \\ \hline
Area (KGE) (logic only) & 114.98                                              & -                                                 & 225                                                     \\ \hline
Area eff. (GOPS/KGE)    & \begin{tabular}[c]{@{}c@{}}20.038\\ \footnotemark[1]20.038\end{tabular}                                              & -                                                 & \begin{tabular}[c]{@{}c@{}}0.286\\ \footnotemark[1]0.644\end{tabular}                                                   \\ \hline
Core power (mW)              & 88.968                                              & 162.4                                             & 0.625                                                   \\ \hline
Power eff. (TOPS/W)     & \begin{tabular}[c]{@{}c@{}}25.9\\ \footnotemark[2]25.9\end{tabular} & \begin{tabular}[c]{@{}c@{}}0.315\\ -\end{tabular} & \begin{tabular}[c]{@{}c@{}}103.14\\ \footnotemark[2]103.14\end{tabular} \\ \hline
\end{tabular}
\label{comparison with other hardware}
\footnotemark[1]{Normalized area efficiency that is scaled to 40nm}

\footnotemark[2]{Normalized power efficiency that is scaled to 40nm and 0.9V}
\end{table}

\subsection{Hardware analysis}
Table.~\ref{comparison with other hardware} shows the performance summary and comparison with other SNN designs. This design is synthesized with TSMC 40nm library using the Synopsys
Design Compiler, achieving peak 2304 GOPS at 500MHz and consumes 88.968 mW for CIFAR-10.  The DRAM access amount is reduced from 1450.172 KB to 938.172 KB with layer fusion, a total reduction of $35.3\%$. Compared to the other reconfigurable design~\cite{narayanan2020spinalflow}, they have lower throughput and power efficiency due to their element wise sparse processing.~\cite{chuang202090nm} is a dedicated five layers SNN ASIC, and thus achieves lower power by eliminating number of memory accesses. But that design cannot be reconfigured for other models and have very low logic area efficiency. 


\section{Conclusion}
In this paper, we integrate the proposed IF-based Batch Normalization into the binary weight spiking neural network to reduce the hardware cost. Our model achieve $90.28\%$ accuracy on CIFAR-10 using only 8 time steps. In addition, the proposed reconfigurable vectorwise accelerator can handle the different models at will, and supports the multi-bit input encoding layer and layer fusion mechanism according to the configuration. Our design can operate at 25.9 TOPS/W under peak efficiency with better power and area efficiency than the previous reconfigurable design and higher flexibility than the fixed network design.

\section*{Acknowledgment}
This work was supported in part by the Ministry of Science and Technology, Taiwan, under Grant 109-2634-F-009-022. 

\bibliographystyle{IEEEtran}
\bibliography{citation.bib}

\begin{thebibliography}{10}
\providecommand{\url}[1]{#1}
\csname url@samestyle\endcsname
\providecommand{\newblock}{\relax}
\providecommand{\bibinfo}[2]{#2}
\providecommand{\BIBentrySTDinterwordspacing}{\spaceskip=0pt\relax}
\providecommand{\BIBentryALTinterwordstretchfactor}{4}
\providecommand{\BIBentryALTinterwordspacing}{\spaceskip=\fontdimen2\font plus
\BIBentryALTinterwordstretchfactor\fontdimen3\font minus
  \fontdimen4\font\relax}
\providecommand{\BIBforeignlanguage}[2]{{%
\expandafter\ifx\csname l@#1\endcsname\relax
\typeout{** WARNING: IEEEtran.bst: No hyphenation pattern has been}%
\typeout{** loaded for the language `#1'. Using the pattern for}%
\typeout{** the default language instead.}%
\else
\language=\csname l@#1\endcsname
\fi
#2}}
\providecommand{\BIBdecl}{\relax}
\BIBdecl

\bibitem{yin2017algorithm}
S.~Yin, S.~K. Venkataramanaiah, G.~K. Chen, R.~Krishnamurthy, Y.~Cao,
  C.~Chakrabarti, and J.-s. Seo, ``Algorithm and hardware design of
  discrete-time spiking neural networks based on back propagation with binary
  activations,'' in \emph{2017 IEEE Biomedical Circuits and Systems Conference
  (BioCAS)}.\hskip 1em plus 0.5em minus 0.4em\relax IEEE, 2017, pp. 1--5.

\bibitem{frenkel2019morphic}
C.~Frenkel, J.-D. Legat, and D.~Bol, ``Morphic: A 65-nm 738k-synapse/mm$^2$
  quad-core binary-weight digital neuromorphic processor with stochastic
  spike-driven online learning,'' \emph{IEEE Transactions on Biomedical
  Circuits and Systems}, vol.~13, no.~5, pp. 999--1010, 2019.

\bibitem{park20197}
J.~Park, J.~Lee, and D.~Jeon, ``A 65nm 236.5 nj/classification neuromorphic
  processor with 7.5\% energy overhead on-chip learning using direct spike-only
  feedback,'' in \emph{2019 IEEE International Solid-State Circuits
  Conference-(ISSCC)}.\hskip 1em plus 0.5em minus 0.4em\relax IEEE, 2019, pp.
  140--142.

\bibitem{chuang202090nm}
P.-Y. Chuang, P.-Y. Tan, C.-W. Wu, and J.-M. Lu, ``A 90nm 103.14 tops/w
  binary-weight spiking neural network cmos asic for real-time object
  classification,'' in \emph{2020 57th ACM/IEEE Design Automation Conference
  (DAC)}.\hskip 1em plus 0.5em minus 0.4em\relax IEEE, 2020, pp. 1--6.

\bibitem{akopyan2015truenorth}
F.~Akopyan, J.~Sawada, A.~Cassidy, R.~Alvarez-Icaza, J.~Arthur, P.~Merolla,
  N.~Imam, Y.~Nakamura, P.~Datta, G.-J. Nam \emph{et~al.}, ``Truenorth: Design
  and tool flow of a 65 mw 1 million neuron programmable neurosynaptic chip,''
  \emph{IEEE transactions on computer-aided design of integrated circuits and
  systems}, vol.~34, no.~10, pp. 1537--1557, 2015.

\bibitem{davies2018loihi}
M.~Davies, N.~Srinivasa, T.-H. Lin, G.~Chinya, Y.~Cao, S.~H. Choday, G.~Dimou,
  P.~Joshi, N.~Imam, S.~Jain \emph{et~al.}, ``Loihi: A neuromorphic manycore
  processor with on-chip learning,'' \emph{IEEE Micro}, vol.~38, no.~1, pp.
  82--99, 2018.

\bibitem{narayanan2020spinalflow}
S.~Narayanan, K.~Taht, R.~Balasubramonian, E.~Giacomin, and P.-E. Gaillardon,
  ``Spinalflow: an architecture and dataflow tailored for spiking neural
  networks,'' in \emph{2020 ACM/IEEE 47th Annual International Symposium on
  Computer Architecture (ISCA)}.\hskip 1em plus 0.5em minus 0.4em\relax IEEE,
  2020, pp. 349--362.

\bibitem{wu2019direct}
Y.~Wu, L.~Deng, G.~Li, J.~Zhu, Y.~Xie, and L.~Shi, ``Direct training for
  spiking neural networks: Faster, larger, better,'' in \emph{Proceedings of
  the AAAI Conference on Artificial Intelligence}, vol.~33, 2019, pp.
  1311--1318.

\bibitem{wu2018spatio}
Y.~Wu, L.~Deng, G.~Li, J.~Zhu, and L.~Shi, ``Spatio-temporal backpropagation
  for training high-performance spiking neural networks,'' \emph{Frontiers in
  neuroscience}, vol.~12, p. 331, 2018.

\bibitem{hubara2016binarized}
I.~Hubara, M.~Courbariaux, D.~Soudry, R.~El-Yaniv, and Y.~Bengio, ``Binarized
  neural networks,'' in \emph{Advances in neural information processing
  systems}, 2016, pp. 4107--4115.

\bibitem{ioffe2015batch}
S.~Ioffe and C.~Szegedy, ``Batch normalization: Accelerating deep network
  training by reducing internal covariate shift,'' \emph{arXiv preprint
  arXiv:1502.03167}, 2015.

\bibitem{yonekawa2017chip}
H.~Yonekawa and H.~Nakahara, ``On-chip memory based binarized convolutional
  deep neural network applying batch normalization free technique on an fpga,''
  in \emph{2017 IEEE International Parallel and Distributed Processing
  Symposium Workshops (IPDPSW)}.\hskip 1em plus 0.5em minus 0.4em\relax IEEE,
  2017, pp. 98--105.

\bibitem{alwani2016fused}
M.~Alwani, H.~Chen, M.~Ferdman, and P.~Milder, ``Fused-layer cnn
  accelerators,'' in \emph{2016 49th Annual IEEE/ACM International Symposium on
  Microarchitecture (MICRO)}.\hskip 1em plus 0.5em minus 0.4em\relax IEEE,
  2016, pp. 1--12.

\bibitem{sengupta2019going}
A.~Sengupta, Y.~Ye, R.~Wang, C.~Liu, and K.~Roy, ``Going deeper in spiking
  neural networks: Vgg and residual architectures,'' \emph{Frontiers in
  neuroscience}, vol.~13, p.~95, 2019.

\bibitem{rathi2020enabling}
N.~Rathi, G.~Srinivasan, P.~Panda, and K.~Roy, ``Enabling deep spiking neural
  networks with hybrid conversion and spike timing dependent backpropagation,''
  \emph{arXiv preprint arXiv:2005.01807}, 2020.

\bibitem{han2020rmp}
B.~Han, G.~Srinivasan, and K.~Roy, ``Rmp-snn: Residual membrane potential
  neuron for enabling deeper high-accuracy and low-latency spiking neural
  network,'' in \emph{Proceedings of the IEEE/CVF Conference on Computer Vision
  and Pattern Recognition}, 2020, pp. 13\,558--13\,567.

\bibitem{wang2020deep}
Y.~Wang, Y.~Xu, R.~Yan, and H.~Tang, ``Deep spiking neural networks with binary
  weights for object recognition,'' \emph{IEEE Transactions on Cognitive and
  Developmental Systems}, 2020.

\end{thebibliography}
\end{document}